\documentclass[11pt]{article}
\usepackage{epsfig}
\def\be{\begin{eqnarray}}\def\ba{\begin{eqnarray}}
\def\ee{\end{eqnarray}}\def\ea{\end{eqnarray}}
\def\ben{\begin{enumerate}}\def\bitem{\begin{itemize}}
\def\een{\end{enumerate}}\def\eitem{\end{itemize}}
\def\no{\nonumber\\}
\def\bi{\bibitem}

\def\Km{K^-}

\def\bfA{{\bf A}}

\def\Tr{{\mbox{Tr}}}

\def\roughly#1{\mathrel{\raise.3ex\hbox{$#1$\kern-.75em%
\lower1ex\hbox{$\sim$}}}}

\def\A0{A_0}
\def\bq{\begin{equation}}
\def\eq{\end{equation}}
\def\la{\langle}\def\ra{\rangle}
\def\Km{K^-}\def\K0{K^0}

\begin{document}

\begin{titlepage}

\begin{center}

 \vskip 1.5cm

{\Large \bf The Effect of Kaon Condensation on Quark-Antiquark Condensate in
Dense Matter}

\vskip 2.3cm

  {\large Youngman Kim$^{(a,b)}$,
 Kuniharu Kubodera$^{(a)}$
Dong-Pil Min$^{(c)}$, Fred Myhrer$^{(a)}$ and Mannque
Rho$^{(c,d)}$ }

\vskip 0.5cm
(a)~{\it Department of Physics and Astronomy,
University of South Carolina, Columbia, South Carolina 29208, USA}

(b) {\it School of Physics, Korea Institute for Advanced Study,
Seoul 130-012, Korea}

(c)   {\it  Department of Phyiscs and Center for Theoretical
Physics, Seoul National University, Seoul 151-742, Korea}

(d)  {\it Service de Physique Th\'eorique, CEA Saclay, 91191
Gif-sur-Yvette  Cedex, France}

\end{center}

\centerline{(\today) }
\vskip 1cm
\vspace{1.0cm plus 0.5cm minus 0.5cm}

\begin{abstract}
Assuming that at sufficiently high densities the constituent quarks
become 
relevant degrees of freedom, we study within the framework of a chiral
 quark model the influence of s-wave $K^-$ condensation on the
 quark-antiquark condensates. 
 We find that, in linear density approximation, the presence of a $K^-$
 condensate quenches
 the $\bar{u}u$ condensate, but that the $\bar{d}d$ condensate remains
 unaffected
 up to the chiral order under consideration.  We discuss the
 implication of
 the suppressed $\bar{u}u$ condensate for flavor-dependent chiral
 symmetry restoration
 in dense matter
\end{abstract}

\end{titlepage}

\newpage

\section{Introduction}

One of the challenging current problems
in nuclear physics is to elucidate
the behavior of nuclear matter under extremely high-density
and/or high-temperature environments.
It is theoretically expected
that, at very high baryon densities
(even at low temperatures),
chiral symmetry is likely to be restored,
and that baryon matter can be
converted into quark matter;
for a review, see e.g. Ref.\cite{birse94}.
Reasonable estimates
also suggest the possible formation
of a kaon condensate
at high densities~\cite{kn,bkrt,tpl94,Leeetal95,chlee,mmttt,mpp}.
The existence of quark matter and/or a
kaon condensate can have important
consequences for the structure of compact stars
and for the cooling behavior
of a remnant star after
supernova explosion and the subsequent
formation of a neutron star.

\vspace{3mm}

Kaon condensation has been studied extensively since Kaplan and
Nelson's seminal work~\cite{kn} appeared in the mid 1980's. In a
tree-order calculation in chiral perturbation theory (ChPT) based
on an $SU(3)_L\times SU(3)_R$ chiral Lagrangian, Kaplan and Nelson
showed that s-wave kaon condensation could occur at a density
around $3\rho_0$, where $\rho_0$ is the normal nuclear density. It
was subsequently pointed out that electrons with high chemical
potential would help speed the condensation process~\cite{bkrt}.
An improved ChPT treatment of kaon condensation that goes beyond
tree-order calculations and that is consistent with the empirical
kaon-nucleon interactions was subsequently 
proposed [3,4,5], and the results indicated
that the critical density $\rho_c^K$ could lie in the range
$2\rho_0\!<\!\rho_c^K\!<\!4\rho_0$; for a review, see {\it e.g.}
Ref.~\cite{chlee}. 
A two-loop calculation of the s-wave pion and kaon self-energies in
nuclear matter was carried out
in Refs.~\cite{kw01,pjm02,BLKetal}. 
These investigations suggested that
the effective pion mass in matter is likely to be relatively stable as a
function of the density. 
The behavior of the in-medium $K^-$
effective mass, $m_K^\star$, 
just below the $K^-p$ threshold is
less clear since it is strongly affected by the non-perturbative
in-medium dynamics of the sub-threshold
$\Lambda(1405)$-resonance~\cite{Leeetal95,lsw92,yabu93}
and by the $\bar{K}N$ coupling to the $\pi\Sigma$ channel, 
see e.g. Ref.\cite{lsw92,yabu93}. 
This topic was discussed in detail in Ref.~\cite{WRW}, 
where Pauli blocking and nucleon-nucleon
short-range correlations were also taken into account in
estimating $m_K^\star$. The effects of kaon-nucleon and
nucleon-nucleon correlations on $m_K^\star$ and kaon condensation
in nuclear medium were studied in Ref.~\cite{chp}, 
according to which these correlations move 
the critical nuclear density for
kaon condensation above $6\rho_0$.
The competition of pion and kaon condensation and 
the phase diagram of a three-flavor Nambu-Jona-Lasinio model 
at finite temperature $T$ and finite quark chemical potentials 
were investigated in Ref.~\cite{bcpr}.
At sufficiently high densities,
strange-quark degrees of freedom may become relevant~\cite{hyper},
and the presence of strange matter can push the onset of kaon
condensation to higher densities and, for some choices of the
input parameters that are hard to pin down, even out of the
physically relevant density regime~\cite{hyper}.

Meanwhile there have recently been interesting developments concerning deeply 
bound kaonic nuclear states and kaonic atoms~\cite{Dka}.  These probes
however pertain 
to the (near-)zero-density environment, and one should keep in mind
that mechanisms extrapolated 
or inferred from the zero-density regime may not be operative at high
densities, in particular,
 in the proximity of chiral phase transition where kaon condensation supposedly takes place. 
For instance, according to a renormalization group flow formalism,  
which is an appropriate framework to use when phase transitions 
are involved, certain terms in the Lagrangian that describe multi-body 
correlations may become ``irrelevant" operators, playing a negligible
role near the transition region~\cite{LRS}.  
This can be the case with the 
four-baryon interaction terms that play a significant role
in the energy-density region where the above-mentioned 
$\Lambda$(1405)-resonance is prominent.  
In other words, it is possible that these four-baryon terms 
are only important at very low 
densities far from the high densities 
required for a phase transition to occur 
(be that the critical density for 
kaon condensation or chiral restoration).
We also note that 
lattice studies of matter in heat bath indicate that the relevant
fermionic degrees of freedom near the chiral transition
temperature are the constituent quarks rather than 
the baryons.
Furthermore, recent developments in holographic dual QCD~\cite{ASY, MMT} at
finite temperture indicate that,
whereas chiral symmetry is definitely broken when confinement exists,
the converse is not necessarily true, and that the chiral symmetry restoration
scale can be much higher than the deconfinement scale.
It has been shown~\cite{MMT} that,
in a regime where chiral symmetry is broken but confinement still persists,
the relevant mass scale is the constituent quark mass, not the current quark mass. 
Although it is perhaps unsafe
to naively apply the same reasoning to the case
at hand (that is, to a high-density case), 
a study based on skyrmion matter~\cite{pseudogap}
indicates that, at high densities, 
basic changes in the matter structure can lead
to in-medium interactions that are significantly different from 
those in free space. An interesting possibility is to adopt the
holographic QCD models with baryons~\cite{SSbaryon} 
to delve into kaon condensate, which, however, will not be pursued in the
present work.

\vspace{3mm}
It is to be noted that the chiral quark model ($\chi$QM) 
enables us to take into account -- at least partially --
the features discussed above, and we consider it illuminating to 
study (possible) kaon condensation in dense matter
in the framework of $\chi$QM.
The relevance of $\chi$QM
in the neighborhood of chiral phase transition was discussed in detail
in Refs.~\cite{riska-brown,BR96,BR01,BR04}.
In analogy to the temperature-induced chiral restoration where $\chi$QM is
invoked~\cite{BLR-star}, the relevant degrees of freedom above
a certain value of density, say, 
$\bar{\rho} > \rho_0$~\footnote{Studies based 
on the effective field theory treatment of nuclear matter indicate
that $\bar{\rho}$ can be somewhat greater than normal matter
density $\rho_0$~\cite{BR04,AHZ}. The precise value of $\bar{\rho}$ which
cannot be pinned down at present is not likely to be important for
our purposes.} 
can be taken to be the constituent quarks whose
masses are generated via the ``dressing" with the ``soft"
component of the gluon field, with the ``hard" component hardly
participating in the process. 
Thus one can think of the
constituent quarks as quasi-particles resulting from highly nonpertubative
vacuum re-structuring caused by the medium (high temperature and/or
density) and hence encapsulating certain aspects of many-body
correlations. 
As a result of this ``dressing",
the constituent quarks, we may call them quasiquarks,
can be expected to interact weakly among themselves and with the 
(hard component of the) gluon field. 
This means that $\chi$QM has the advantage of
providing a systematic chiral power counting for the interactions
of the constituent quarks with the Goldstone bosons and, in
addition, allowing a weak-coupling expansion for interactions with
the gluons~\cite{mg}. 
This aspect gives a justification
for us to ignore gluonic contributions in calculating 
the effective potential (see below).
This approach to kaon condensation anchored on a chiral quark
Lagrangian valid above the density $\bar{\rho}$ is consistent with
-- and complements -- the top-down approach~\cite{BLR05} based
on expansion around the vector manifestation fixed point
(which coincides with the chiral restoration point)
of the Harada-Yamawaki hidden local symmetry theory~\cite{HY:PR}.

\vspace{3mm}

In the present work we study $\Km$ condensation 
in the framework of $\chi$QM~\cite{mg}, and 
undertake the first investigation of 
how and to what extent a possible kaon condensate 
distorts the Fermi seas of the quasi-quarks 
and what influences this distortion can have 
on chiral symmetry restoration in dense matter. 
The issues investigated here are important  
in connection with phase transitions leading to color superconductivity;
the behavior of the transition in the chiral quark picture could be quite
different from that of the standard scenario, where the transition is 
presumed to occur from a Fermi liquid state. 
In the present exploratory study, 
we describe quark matter as a free Fermi gas of quasiquarks. 
This treatment does not explicitly take into account the strong
correlations believed to be present at 
ordinary (low) nuclear matter density,
at which three quarks are clustered into color-singlet
nucleons that are spatially separated by strong short-range
(nucleon-nucleon) repulsion. 
However, there exists the expectation that, unlike the baryons, 
the constituent quarks are not susceptible 
to strong short-range correlations~\cite{BLR-star},
and hence the neglect of correlation effects is likely 
to be a less serious problem in the $\chi$QM approach than
in the baryonic picture.
Although this problem warrants further examinations,
we limit ourselves here to the Fermi gas model
and investigate (within the confine of this model)
the consequences of the $\chi$QM Lagrangian -- 
assumed to be valid above a few
times the nuclear matter density -- on kaon condensation 
and also the effects of kaon condensation (if it occurs)
for the quark-antiquark condensate.

\vspace{3mm}

The paper is organized as follows. In section 2 we first  give a brief
recapitulation of $\chi$QM, and then we demonstrate that,
within the framework of linear density approximation, the use of
$\chi$QM for describing $\Km$ condensation essentially reproduces
the results obtained both in heavy-baryon chiral perturbation
theory (HBChPT)~\cite{chlee} and in a formalism 
that describes fluctuations around the vector
manifestation fixed point of hidden local symmetry
theory~\cite{BLR05}.
In section 3
we examine the effects of $\Km$ condensation on
the quark-antiquark
condensate $\la\bar q q\ra$
in the framework of $\chi$QM
and in HBChPT.
Section 4 is dedicated to summary.
In Appendix A, we present a brief discussion on
the power counting rules in $\chi$QM.

\section{Kaon condensation in the chiral quark model}
The chiral quark model ($\chi$QM)
we employ here is defined by the
Lagrangian~\cite{mg} \ba {\mathcal L}&=&{\mathcal L}_0 +{\mathcal
L_M}+{\mathcal L}_{m_\phi},\label{ChQM} \ea where the
chiral-symmetry invariant part ${\mathcal L}_0$ is given by
 \ba
{\mathcal L}_0
&=&{\bar\psi}(i {\not\!\! D}+{\not\! V})\psi +
g_{A}{\bar \psi}{\not\!\! A}\gamma_5\psi
               - M_0 {\bar \psi}\psi \no
               &&+ \frac{1}{4}f_\pi^2
\Tr(\partial^{\mu}\Sigma^{\dag}\partial_{\mu}\Sigma)
               - \frac{1}{2}\Tr(G^{\mu\nu}G_{\mu\nu}) + \ldots
\label{lag}
\ea
where $G^{\mu\nu}$ is the QCD field tensor.  
The covariant derivative is defined by
\ba
D_\mu=\partial_\mu+i g G_\mu;~~~ G_\mu=G_\mu^a T^a;
\label{CoD}
\ea
where $G^a_\mu$ is the gluon field with $a$ $= 1, \cdots, 8$ and
\ba
V_\mu&=&{i \over
2}(\xi^{\dagger}\partial_\mu\xi+\xi\partial_\mu\xi^{\dagger});~~~
A_\mu={i \over
2}(\xi^{\dagger}\partial_\mu\xi-\xi\partial_\mu\xi^{\dagger});
\no
\psi &=&
 \left( \begin{array}{c}
u\\
d\\
s
 \end{array} \right);~~~
\xi=e^{(i \Pi/f_\pi)};~~~
\Sigma=\xi\xi\nonumber
\ea
with
\ba
\Pi={1 \over\sqrt{2}}\left( \begin{array}{ccc}
                            \sqrt{1 \over 2}\pi^0+\sqrt{1 \over 6}\eta&
\pi^+& K^+ \\
                            \pi^-&-\sqrt{1\over 2}\pi^0+\sqrt{1\over
6}\eta& K^0 \\
                            K^- & \bar{K}^0& -{2 \over \sqrt{6}}\eta
\end{array} \right).
\label{L1}
\ea
Here $ f_\pi\simeq 93~ {\rm MeV}$, and
$M_0\approx 350$ MeV
denotes the part of the constituent
quark mass generated by spontaneous chiral symmetry breaking which
constitutes the bulk of the effective quark mass. The chiral
symmetry-breaking term ${\mathcal L_M}$ is given by
 \ba
{\mathcal L_M}= -\frac{1}{2}c_1\bar\psi(\xi^\dagger {\mathcal
M}\xi^\dagger +\xi {\mathcal M} \xi)\psi \label{L2}
\ea
where
\ba
 {\mathcal M}=
\left(
\begin{array}{ccc}
 m_u& 0& 0 \\
0& m_d& 0 \\
0& 0 & m_s
    \end{array} \right),\nonumber
\ea
and $c_1\approx 1$ \cite{mg}, which is fixed from the mass
difference ($\sim 150 ~{\rm  MeV}$) between the $s$ quark and the
$u$ (or $d$) quark.
We will assume $c_1=1$ in this work.
${\mathcal L}_{m_\phi}$ in Eq. (\ref{ChQM}), which is responsible
for the finite Goldstone boson masses, takes the form \ba
{\mathcal L}_{m_\phi}= \frac{1}{2}f_\pi^2 {\rm Tr} (\mu {\mathcal
M} \Sigma^\dagger ) +h.c.~,\label{lma} \ea where $\mu$ is a
parameter with the dimension of mass.

\vspace{3mm}

$\chi$QM offers a systematic chiral power counting expansion
in describing the interactions of constituent quarks with
Goldstone bosons (summarized in Appendix A).
To estimate higher-loop corrections,
we need to set up in-medium power counting rules
in $\chi$QM.\footnote{
In-medium chiral counting rules
in the hadronic picture
have been discussed at length
in Refs.~\cite{MOW}.}
To introduce in-medium counting rules applicable
to dense matter,
we assume that $p\sim k_F$,
where $p$ is a typical momentum scale and
$k_F$ is the quark Fermi momentum.
Since the $u$- and $d$ quark Fermi momentum
at $ 4\rho_0$ (symmetric matter)
is about $430$ MeV,
the expansion parameter related to the Fermi momentum is
$k_F/\Lambda\sim 0.43$ ($\Lambda=1$ GeV).
With the assumption $p\sim k_F$,
it is easy to establish that the
in-medium counting rules
agree with the free-space counting rules;
see Eq.(\ref{ct}) in Appendix A.
It is to be noted that in ChPT
involving the nucleons,
in-medium chiral counting contains
a subtle issue related
to the ``heavy" nucleon mass~\cite{MOW}.
No such issue arises in $\chi$QM,
since the constituent quark masses are of the same order
as the typical momentum scale $p$.
In addition, $\chi$QM allows for
a perturbative expansion
of the constituent-quark gluon interaction
(with $\alpha_s\approx 0.28$)~\cite{mg}.
This scheme contains no free
parameters to the chiral order we will consider in this work
and to lowest order in $\alpha_s$.
Those terms in the
Lagrangian Eq. (\ref{ChQM}) which are relevant for
our discussion of kaon condensation in quark matter
can be written as
\ba
{\mathcal L}_K&=& \frac{i}{4f_\pi^2} [\bar u
(K^+\not\!\partial K^- -(\not\!\partial K^+)K^- )u + \bar s
(K^-\not\!\partial K^+ -(\not\!\partial K^-)K^+ )s ]
\label{LL1}\no
&+& \frac{1}{2f_\pi^2}(m_u +m_s)
[\bar u K^+K^- u+ \bar s K^- K^+
s] \; .
\label{LL2}
\ea

%
\begin{figure}

\centerline{\epsfig{file=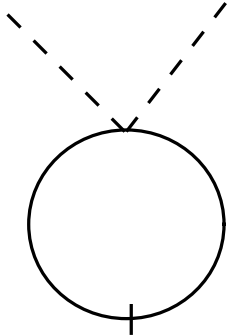,width=3.0cm}} \caption{\small
The kaon self-energy contribution in quark matter at
lowest chiral order.
The dashed line stands for the kaon,
and the solid line for the quark.
The quark propagator line marked with ``$|$''
denotes the density-dependent part of the
quark propagator (at zero temperature),
$-2\pi\delta (k^2-M_f^2)
\theta (k_0)N_F (k_0)|_{T\rightarrow 0}$,
where $N_F(k_0)$ is
the Fermi-Dirac distribution function
and $M_f$ is the mass of the constituent quark
of flavor $f$. }
\label{kaon}
\end{figure}
\begin{figure}
\centerline{\epsfig{file=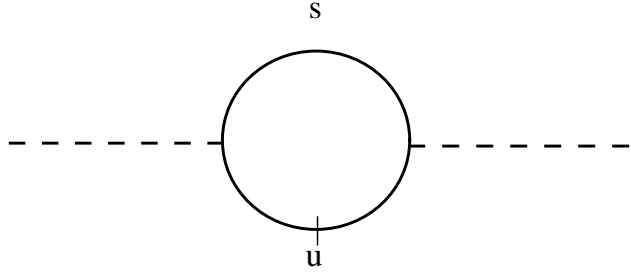,width=8.3cm}}
\caption{\small Kaon self-energy.
The dashed lines stand for kaons and
  the solid lines for quarks.
  The symbol ``$|$'' denotes the
  density-dependent part of
  the quark propagator.}
\label{pkc}
\end{figure}

In order to establish a basis for our work,
we first describe kaon condensation in the framework of
$\chi$QM with the purpose of comparing
our results with those obtained in HBChPT~\cite{chlee}.
We calculate the $K^-$ self-energy using the Lagrangian
${\mathcal L}_K$ in Eq.(\ref{LL2}).
The lowest-order contributions
come from the graphs shown in Fig. \ref{kaon} and
Fig. \ref{pkc}. 
As we will show below, however,
the contribution from
Fig. \ref{pkc} is negligible compared
to the one from Fig. \ref{kaon}.
Higher chiral-order contributions
will not be discussed in this work.

We consider kaons in medium that solely consists
of up and down quarks with no strange-quarks in the
Fermi sea.
Our quark matter is assumed to be
symmetric with respect to the u- and d-quarks,
so the quark densities in the present case
are characterized by
$\rho_u=\rho_d\equiv\rho_q$ and $\rho_s=0$,
while the baryon density $\rho_B$ is given by
$\rho_B=\frac{2}{3}\rho_q$.
To establish connection with the previous works, 
we will demonstrate 
that, within the framework of the linear density
approximation, the $\chi$QM approach to kaon condensation leads to
results similar to those obtained in
HBChPT\cite{bkrt,tpl94,Leeetal95,chlee} and those obtained
in an expansion around the vector manifestation fixed point~\cite{BLR05}.
%
To compare with the HBChPT calculations, we need to
use non-relativistic approximation
in evaluating the kaon self-energy in Fig. \ref{kaon}.
Since the constituent quark mass,
$M_0 \approx 350$ MeV, is smaller than the chiral scale,
$\Lambda \simeq 1$ GeV, and is of the order of
$k_F$, this non-relativistic
approximation might not be very reliable.
It is used here $only$ for the sake of comparison
with previous works.
Using the lagrangian in Eq.(\ref{LL2}), we find
 \ba
-i\Sigma_K (q_0) = i\,
[\,\frac{3}{4}\, \frac{(m_u+m_s)}{f_\pi^2} +\frac{3}{4}
\frac{q_0}{f_\pi^2}\,]\,\rho_B
 \ea
The in-medium kaon mass $m_K^{\star}$
is then obtained by solving the dispersion equation
\ba
m_K^{\star 2} =m_K^2 + \Sigma_K (q_0=m_K^{\star }),
\label{Dispersion1}
 \ea
where $m_K \approx 500$ MeV is the free kaon mass.
We define $x=\frac{m_K^\star}{m_K}$
and $c=\rho_B/\rho_0$,
where $\rho_0 = 0.17$ fm$^{-3}$ is
the normal nuclear matter density,
to rewrite the dispersion equation as
\ba x^2 +0.24cx +0.12c-1 =0\, , \label{Dispersion2}
\ea
where we have used for the current quark masses
$m_u\approx 6$ MeV,
$m_s\approx 240$ MeV as in \cite{chlee}.
Solving this equation for typical values of $c$,
we arrive at
\ba
c=1&:&m_{K^-}^\star\approx 410\  {\rm MeV} \no
c=2&:&m_{K^-}^\star\approx 330\  {\rm MeV} \no
c=3&:&m_{K^-}^\star\approx 260\  {\rm MeV} \no
c=4&:&m_{K^-}^\star\approx 193\  {\rm MeV}.
\label{kmassNR}
\ea
We note that the HBChPT calculation
in Ref.~\cite{chlee} finds
$m_{K^-}^\star\approx 360$ MeV for $c=1$.
Since the electron chemical potential $\mu_e$
is known to have values
$200\,{\rm MeV}\!\!<\!\!\mu_e \!\!<\!\!300\, {\rm MeV}$
in the density range
$\rho_B=(2-4)\rho_0$~\cite{tpl94},
one may infer that kaon condensation takes
place around $\rho_B\sim (3-4)\rho_0$.
In the above we ignored the contribution
from the diagram in Fig.\ref{pkc}
and furthermore, to facilitate comparison with HBChPT,
we used the heavy-fermion approximation for the quark propagator.
We now examine whether the contribution of Fig. \ref{pkc} to
s-wave kaon condensation is indeed negligible. We also study the
consequences of the relativistic treatment of the quark
propagator.
%
%
To this end, we calculate {\it relativistically}
the contributions of the diagrams
in  Figs.~\ref{kaon} and \ref{pkc}.
For SU(2) symmetric matter,
the kaon self-energy, $ -i\Sigma_K^{ex} (q_0)$,
corresponding to Fig.~\ref{pkc} is
given by
\ba
\Sigma_K^{ex} (q_0)=
\frac{g_A^2}{2\pi^2f_\pi^2}\; q_0^2 \;\int_0^{k_F}
d\bar k \left(\frac{\bar k^2}{k_0}\right)\left(
\frac{2k_0^2-M_u^2+q_0k_0-M_uM_s}
{q_0^2+2q_0k_0+\Delta M^2}\right),
\ea
where $\bar k=|\vec k|$, $k_0=\sqrt{\bar k^2+M_u^2}$
and $\Delta M^2=M_u^2-M_s^2$,
with $M_u$ and $M_s$ being the masses
of the constituent $u$- and $s$-quarks, respectively.
We again solve the dispersion equation,
$m_K^{\star 2} =m_K^2 +\Sigma_K(q_0=m_K^{\star})$,
with the kaon self-energy $\Sigma_K$
which includes the contributions
of the diagrams in Fig.~\ref{kaon} and Fig.~\ref{pkc}.
The results are given in Eq.(\ref{rr}).
For each row, the number that appears to the left (right)
of the arrow is the result that excludes (includes)
the contribution of Fig. \ref{pkc}.
\ba
c=1&:&m_{K^-}^\star\approx
415\, {\rm MeV}\rightarrow 424\, {\rm MeV}\no
c=2&:&m_{K^-}^\star\approx
341\, {\rm MeV}\rightarrow 361\, {\rm MeV}\no
c=3&:&m_{K^-}^\star\approx
278\, {\rm MeV}\rightarrow 313\, {\rm MeV}\no
c=4&:&m_{K^-}^\star\approx
225\,  {\rm MeV}\rightarrow 283\, {\rm MeV}\; .
\label{rr}
\ea
Comparison of the results in Eq.(\ref{kmassNR})
with those in Eq.(\ref{rr})
reveals that the relativistic corrections
are small,
ranging from about 1\% to $\sim 17$\%.
Treating the diagram in Fig.\ref{kaon} relativistically
increases the value of $m_{K^-}^\star$
compared to the value of the
non-relativistic treatment,
especially for the higher densities
$\rho_{\rm B}\sim (3-4)\rho_0$.
As mentioned, the change
in the value $m_{K^-}^\star$ that occurs
as we add the contribution of Fig.~\ref{pkc}
to that of Fig.~\ref{kaon}
is indicated by the arrow in Eq.(\ref{rr}).
It is seen that
the effect of the diagram in Fig.\ref{pkc}
is a little bit bigger than the relativistic corrections, and it could
push up the critical density for 
 $\Km$ condensation slightly. 

\section{Kaon Condensation and Quark-antiquark Condensate.}
%
As discussed above,
it is possible that a
$\Km$ condensate exists at high matter densities. 
To confirm the existence of  $\Km$ condensation 
in the framework of $\chi$QM, we need to consider the higher
order corrections in a systematic way.
In this section, however, we assume that  $\Km$ condensation 
is realized in dense nuclear matter.
Now, if a kaon condensate is formed,
the values of various quantities that characterize
the physical condition of dense matter can also change.
In this subsection we study
the influence of $\Km$ condensation (if it occurs)
on the quark-antiquark condensate $\la \bar qq\ra_{\rho_B}$
in dense matter. The purpose is to
examine the effects of a postulated kaon condensate
on the possible restoration of chiral symmetry in dense matter.
We will continue to consider
symmetric matter with no strangeness specified by
$\rho_u=\rho_d\equiv\rho_Q=(3/2)\rho_B$ and $\rho_s=0$.

\subsection{Quark-antiquark condensate in the chiral quark model}
We begin with a brief summary of the quark-antiquark condensate at
low density. From the energy density of quark matter calculated in
the standard manner, we can obtain the in-medium quark condensate
with the help of the Hellmann-Feynman theorem and the
Gell-Mann-Oakes-Renner (GMOR) relation, $2m_q\!<\!\bar q
q\!>_{vac}=-m_\pi^2 f_\pi^2$, where $\la {\mathcal O}\ra_{vac}$
represents the vacuum expectation value of the operator ${\mathcal
O}$. The resulting expression for the quark condensate in medium
is 
\ba <\!\bar q q\!>_{\rho_B}= <\!\bar q q\!>_{vac} +
\frac{1}{2}\frac{d\tilde\epsilon}{d m_q}\,\label{qq} 
\ea 
where
$\tilde\epsilon$ is the energy density, and $m_q$ is the current quark
mass, $m_q\equiv (m_u+m_d)/2$. We have assumed here that SU(2)
isospin symmetry is conserved and defined $\bar q q\equiv(\bar u
u+\bar dd)/2$. At low density, Eq. (\ref{qq}) leads to the model
independent result
 \ba \frac{<\!\bar q q\!>_{\rho_B}}
 {<\!\bar q q\!>_{vac}} \simeq
1-\frac{\sigma_N}{m_\pi^2f_\pi^2}\rho_B~,\label{mir}
\ea
where
$\sigma_N$ is the nucleon sigma-term,
see, {\it e.g.}, Ref.\cite{cfg} for details.
The evaluation of corrections to Eq.(\ref{mir})
requires model calculations.\footnote{
In one such model calculation~\cite{cfg}
based on the hadronic picture with
the parameters $M_N$, $m_\pi$ and $g_{\pi NN}$
characterizing the Lagrangian,
Eq.~(\ref{qq}) is evaluated from
\ba
<\!\bar q q\!>_{\rho_N}= <\!\bar q q\!>_{vac} + \frac{1}{2}
[\frac{d M_N}{d m_q}\frac{\partial\tilde \epsilon}{\partial M_N}+
\frac{d m_\pi}{d m_q}\frac{\partial \tilde\epsilon}{\partial m_\pi}+
\frac{d g_{\pi NN}}{d m_q}\frac{\partial\tilde \epsilon}
{\partial g_{\pi NN}}].
\label{qqN}
\ea
}

\vspace{3mm}

We now consider the effects a kaon condensate
can have on the quark-antiquark condensate in matter.
We will find that the existence of a $\Km$ condensate
leads to asymmetry between
$\la \bar u u\ra_{\rho_B} $
and $\la \bar dd\ra_{\rho_B}$.
In the following therefore we treat
$\la \bar uu\ra_{\rho_B}$ and
$\la\bar dd\ra_{\rho_B}$ as independent quantities.
In $\chi$QM characterized by
the Lagrangian Eq.(\ref{lag}), we
are inspired by \cite{cfg} to write:
\ba
<\!\bar q_fq_f\!>_{\rho_B}= <\!\bar q_fq_f\!>_{vac} &+&\frac{1}{2}
[\frac{\partial \tilde\epsilon}{\partial m_f}+
\frac{d M_f}{d m_f}\frac{\partial \tilde\epsilon}{\partial M_f}+
\frac{d m_\phi^2}{d m_f}
\frac{\partial \tilde\epsilon}{\partial m_\phi^2}
]\,
\label{qqQ}
\ea
where $\phi=\pi,~\eta,$  or $K$,
\footnote{In the last term of Eq.~(\ref{qqQ}), we need to sum over
all Goldstone bosons such as contributions from pions~\cite{LKetal}.
In the present work, however, we only include the contribution 
from kaon condensate to highlight 
its specific effects on quark-antiquark condensate.} 
while $f=u,d$, and $\la \bar q_uq_u\ra =\la\bar uu\ra$,
  $\la \bar q_dq_d\ra =\la\bar dd\ra$;
$\tilde\epsilon$ is the energy density,
and $m_\phi$ is the finite Goldstone boson mass
generated by explicit chiral symmetry breaking.
Here we assume that the
coupling constants, $g_A$, $g$ and $f_\pi$ defined in
Eqs.(\ref{lag}) and (\ref{CoD}),
do not depend on the current quark mass,
e.g. $dg/dm_f=0$.
If we assume
the presence of a $K^-$ condensate in $\chi$QM,
a consideration similar to
the one used in Ref.~\cite{bkrt} leads to
the following energy density:
\ba
&&\tilde\epsilon =\frac{3}{4\pi^2}\sum_{f=u,d} [
k_F^f(k_F^{f2}+M_f^2)^{3/2}-
\frac{1}{2}M_f^2k_F^f\sqrt{k_F^{f2}+M_f^2}- \frac{1}{2}
M_f^4\ln\frac{ k_F^f+ \sqrt{k_F^{f2}+M_f^2}}{M_f}]\no
&&\qquad\qquad-\frac{1}{2}f_\pi^2\mu_e^2\sin^2\theta+
2m_K^2f_\pi^2\sin^2\frac{\theta}{2} -\mu_e
\rho_u\sin^2\frac{\theta}{2} -m^\prime\rho_s^u
\sin^2\frac{\theta}{2},
\label{edcq}
\ea
where $\rho_s^u$ is the
scalar density of the $u$ quark and
$m^\prime=m_u+m_s$.
Furthermore, the ``chiral angle" $\theta$
in V-spin space is given by
$\theta\equiv \sqrt{2}v/f_\pi$,
where $v$ is the magnitude of the $\Km$ condensate.
%
%
Considering the energy
density in $\chi$QM given in
Eq.(\ref{edcq}),
we write Eq. (\ref{qqQ}) in terms of
$\la\bar uu\ra$ and $\la\bar dd\ra$:
\ba
\la\bar uu\ra_{\rho_B}&=& \la\bar uu\ra_{vac} +\frac{1}{2}\left[
\frac{\partial \tilde\epsilon}{\partial m_u}+
\frac{d M_u}{d m_u}\frac{\partial \tilde\epsilon}{\partial M_u}+
\frac{d m_K^2}{d m_u}\frac{\partial \tilde\epsilon}{\partial
  m_K^2}\right]
\label{uuQ}\\
\la\bar dd\ra_{\rho_B}&=& \la\bar dd\ra_{vac} +\frac{1}{2}\left[
\frac{d M_d}{d m_d}\frac{\partial \tilde\epsilon}{\partial M_d}+
\frac{d m_K^2}{d m_d}\frac{\partial \tilde\epsilon}{\partial
  m_K^2}\right]\;.
\label{ddQ}
\ea
In the expression for the energy density
Eq.(\ref{edcq}), $\la\bar dd\ra_{\rho_B}$ is independent
of the $\Km$ condensate amplitude $\theta$, which will be shown
explicitly below.
We therefore focus on $\la\bar uu\ra_{\rho_B}$.
In order to evaluate Eq.(\ref{uuQ}) and
deduce the density dependence of
the quark-antiquark condensate,
we require specific information about
$\frac{d M_f}{d m_f}$ and $\frac{d m_\phi^2}{d m_f}$.

\vspace{3mm}

To determine $dM_f/dm_f$, we compare Eq.(\ref{qqQ})
with the model-independent result in Eq.(\ref{mir})
at low density.
Ignoring interactions among the constituent quarks
and the contribution from constituent quark kinetic energy,
which seems to be reasonable at
low density~\cite{cfg},
we obtain the following energy density in $\chi$QM,
\ba
\tilde\epsilon=M_u\rho_u +  M_d\rho_d,\label{eq:E0}
\ea
where $\rho_f\equiv
(k_F^q)^3/\pi^2$.
For symmetric matter we obtain
\ba
\frac{\la \bar u u\ra_{\rho_B}}{\la \bar u u\ra_{vac}}
=1-
\frac{3\sigma_u}{2m_\pi^2f_\pi^2}\rho_B ,
\label{mir-1}
\ea
where $\sigma_u \equiv m_udM_u/dm_u$,
and we have used the GMOR relation.
Comparing  Eq. (\ref{mir-1})  with Eq. (\ref{mir}),
we arrive at
$\sigma_u=2\sigma_N/3$. For the $d$ quark we obtain $\sigma_d=2\sigma_N/3$.
\footnote{As 
mentioned in the introduction, the precise value of 
$\bar\rho$ is not well determined at present, but this is not so
important in the present work. What matters here is the fact
that the relevant degrees of freedom 
near chiral symmetry restoration are the quasiquarks 
rather than the baryons. }
To proceed we now assume that this relation is also
valid up to a few times the normal nuclear matter
density $\rho_0$.
For a numerical estimate
we use the value $\sigma_N= 30$ MeV~\cite{sainio}.

\vspace{3mm}

Next we evaluate $dm_K^2/dm_f$.
 {}The mass term in Eq. (\ref{lma})
 leads to
$m_K^2=\mu (m_u+m_s)$, from which we obtain~\cite{cfg} \footnote{A
comment is in order here on how the nonlinear quark mass
dependence of the Goldstone boson mass affects Eq. (\ref{mkm}).
The following terms give $m_q^2$ corrections to the Goldstone
boson masses, \ba \frac{f_\pi^2}{4}c_1{\Tr} ({\mathcal
M^\dagger}\Sigma ) {\Tr} ({\mathcal M^\dagger}\Sigma )
+\frac{f_\pi^2}{4}c_2 {\Tr} ({\mathcal M^\dagger} \Sigma {\mathcal
M^\dagger}\Sigma ) +\frac{f_\pi^2}{4}c_3 {\Tr} ({\mathcal
M}^\dagger\Sigma ) {\Tr} ({\mathcal M}\Sigma^\dagger )
+h.c.,\nonumber \ea where $c_1$, $c_2$ and $c_3$ are dimensionless
parameters assumed to be of the order of 1. These terms lead to
\ba \frac{dm_K^2}{dm_u}=\frac{m_K^2}{m_u+m_s}
\{1+2(c_1+c_2+c_3)\frac{(m_u+m_s)^2}{m_K^2}+
(c_1+c_3)\frac{m_d(m_u+m_s)}{m_K^2}\}. \nonumber \ea These
corrections are suppressed by a factor $(m_q/m_K)^2$ compared to
the leading term in Eq. (\ref{mkm}). } 
\ba
\frac{dm_K^2}{dm_u}=\frac{m_K^2}{m_u+m_s} \; \; \; \; (= \mu) \; ,~
 \frac{dm_K^2}{dm_d}=0\; .
\label{mkm} 
\ea

The amplitude of the kaon condensate
$\theta$ can be determined by extremizing Eq. (\ref{edcq})
with respect to $\theta$.
The result is
\ba
\cos\theta=\frac{1}{f_\pi^2\mu_e^2} (m_K^2f_\pi^2
-0.5\mu_e\rho_u-0.5m^\prime\rho_s^u). \label{cost}
\ea
Now we consider in-medium $\la \bar dd\ra$ in a kaon-condensed phase.
{}From Eq. (\ref{ddQ}) and Eq. (\ref{mkm}), we obtain
\ba
\la\bar dd\ra_{\rho_B}&=& \la\bar dd\ra_{vac} +\frac{1}{2}
\frac{d M_d}{d m_d}\frac{\partial \tilde\epsilon}{\partial M_d}\;.
\ea
This equation together with the energy density in Eq. (\ref{edcq})
show that the $\Km$ condensate will not affect the $\bar dd$ condensate.  
To see the effects of kaon condensation on the
$u$ quark condensate $\la \bar uu\ra$,
we rewrite Eq. (\ref{uuQ}) as
\ba
R\equiv
\frac{<\bar uu >_{\rho_B}}{<\bar uu>_{vac}}
=1 -\frac{m_u}{m_\pi^2f_\pi^2}
[\frac{\partial \tilde\epsilon}{\partial m_u}+
\frac{\sigma_u}{m_u} \frac{\partial \tilde\epsilon}{\partial
M_u}+ \frac{m_{K}^2}{m^\prime} \frac{\partial
\tilde\epsilon}{\partial
  m_K^2}]\ ,
\label{qcnm-1}
\ea
where we have used Eq.(\ref{mkm}) and
have taken $m_u= m_q$.
The above expressions lead to
\ba
\frac{\partial \tilde\epsilon}{\partial m_u}
&=&f_\pi^2\mu_e^2\cos\theta
\frac{\partial \cos\theta}{\partial m_u}
-m_K^2f_\pi^2\frac{\partial \cos\theta}{\partial m_u}
\no
&&+\frac{1}{2}\mu_e\rho_u \frac{\partial \cos\theta}{\partial m_u}
-\rho_s^u\sin^2\frac{\theta}{2}
+\frac{1}{2}m^\prime  \rho_s^u\frac{\partial \cos\theta}{\partial
  m_u},\no
\frac{\partial \tilde\epsilon}{\partial M_u}&=&
\frac{3}{2\pi^2}( M_uk_F^u\sqrt{k_F^{u2}+M_u^2}-
M_u^3\log\frac{ k_F^u+\sqrt{k_F^{u2}+M_u^2}}{M_u})\no
&&+f_\pi^2\mu_e^2\cos\theta\frac{\partial \cos\theta}{\partial M_u}
-m_K^2f_\pi^2\frac{\partial \cos\theta}{\partial M_u}
+\frac{1}{2}\mu_e\rho_u\frac{\partial \cos\theta}{\partial
  M_u}\no
&&+\frac{1}{2}m^\prime\rho_s^u\frac{\partial \cos\theta}{\partial
  M_u}
-m^\prime \sin^2\frac{\theta}{2}\frac{\partial\rho_s^u}{\partial M_u}
\no
\frac{\partial \tilde\epsilon}{\partial m_K^2}&=&
f_\pi^2(1-\frac{m_K^2}{\mu_e^2}) +\frac{1}{2\mu_e}\rho_u
+\frac{m^\prime}{2\mu_e^2}\rho_s^u\ ,
\ea
with
\ba
\frac{\partial \cos\theta}{\partial
  m_u}&=&-\frac{\rho_s^u}{2f_\pi^2\mu_e^2} ,\no
\frac{\partial \cos\theta}{\partial
  M_u}&=&-\frac{m^\prime}{2f_\pi^2\mu_e^2}
\frac{\partial\rho_s^u}{\partial M_u}\no
\frac{\partial\rho_s^u}{\partial M_u}&=&
\frac{3}{2\pi^2}[ k_F^u\sqrt{k_F^{u2}+M_u^2}+\frac{2M_u^2k_F^u}{\sqrt{k_F^{u2}+M_u^2}}
-3M_u^2\ln\frac{ k_F^u+\sqrt{k_F^{u2}+M_u^2}}{M_u}] .
\ea
If we use for illustrative purposes
$\mu_e=270$ MeV, $\rho_B=3.3\rho_0$,
$\sigma_N= 30~{\rm MeV}$ and
$\cos\theta=0.89$ in Eq.~(\ref{qcnm-1}),
we arrive at ${\rm R} \sim 0.35$.
This should be compared with ${\rm R}\sim 0.39$
that would result if there is no kaon condensation
($\theta=0$).
%
\begin{figure}
\begin{center}
\epsfxsize = 10cm
\ \epsfbox{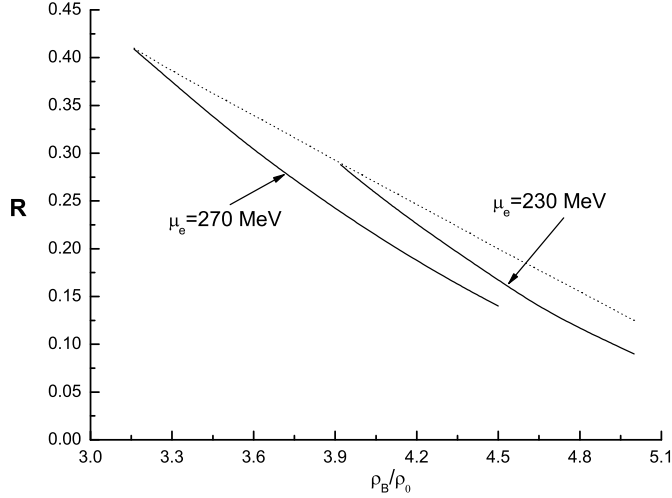}
\end{center}
\vskip -0.7cm
\caption[]{$R \equiv \,
<\!\bar uu \!>_{\rho_B}/<\!\bar uu\!>_{vac}$
calculated with (solid line) and without (dotted line)
kaon condensation taken into account.
The results are given
for two different values of the electron chemical
potential, $\mu_e$= 230 and 270 MeV.
}\label{mue273}
\end{figure}
%
The results for $\rm {R}$ are also shown in
Fig.~\ref{mue273} for
typical cases of $\sigma_N=30$~MeV and $\mu_e$ = 230 and 270
MeV.\footnote{For simplicity, the chemical potential
$\mu_e$ is treated here as an external parameter, although in a
full treatment it should be determined self-consistently, see,
{\it e.g.}, Ref.~\cite{tpl94, kubodera92}.}
Fig.~\ref{mue273} demonstrates
that the presence of a kaon condensate leads to a
faster decrease of $\la \bar uu\ra$ with increasing density.
By contrast, the $d$ quark condensate, $\la \bar dd \ra$, is not
affected by the presence of $K^-$ condensation to
the lowest order in density under
consideration.
We therefore expect that in the kaon condensed
phase $\la \bar uu\ra/\la \bar dd\ra\rightarrow 0$ as $\rho_B$
increases. It is generally expected that chiral symmetry
restoration characterized by $\la \bar uu\ra=\la \bar dd\ra=0$
occurs in dense matter.\footnote{Another order parameter for
chiral symmetry breaking is the pion decay constant $f_\pi$ in the
chiral limit $m_q=0$. Even if $\la \bar uu\ra=\la \bar dd\ra=0$,
chiral symmetry might be still broken, as long as $f_\pi\ne
0$~\cite{MOW}. Such a pseudo-gap phenomenon is observed in dense
skyrmion crystal matter~\cite{pseudogap}.} Our results indicate,
however, that at densities between $\rho_c^K$ and $\rho_c^{\chi
SR}$ ($\rho_c^K <\rho_B< \rho_c^{\chi SR}$), kaon condensation may
lead to a phase characterized by $\la\bar u u\ra \simeq 0$, $\la
\bar dd\ra\neq 0$, which represents partial ``flavor-dependent
restoration" of chiral symmetry.

Finally, we observe that a $\Km$ condensate
gives rise to difference between
the masses of the constituent $u$ and $d$ quarks.
It is easy to show
\ba
M_u^\star&=& M_0 +m_u-\frac{m^\prime}{2f_\pi^2}  v^2\no
M_d^\star&=& M_0 +m_d \; \; \; (=M_d) \ .
\label{isob}
\ea
Thus the presence of a $\Km$ condensate induces
additional $SU(2)$-isospin symmetry breaking
on top of the small explicit isospin breaking due
to $m_u-m_d\neq 0$.

\subsection{Quark-antiquark condensate in HBChPT. }
%
The results shown in previous subsection
were obtained with the use of the energy density
calculated in $\chi$QM
{\it without} taking into account
the beta-equilibrium condition
or the charge-neutrality condition.
In order to check whether
the imposition of these constraints
affects our results,
we evaluate in this section
the quark-antiquark condensates
in a kaon condensed phase using the
energy density that has been calculated
in HBChPT with the beta-equilibrium
and charge-neutrality conditions
taken into account~\cite{bkrt, tpl94, kubodera92}.
The energy density
$\epsilon_{\rm HB}$ in the $K^-$ condensed phase
obtained from HBChPT is~\cite{bkrt, tpl94, kubodera92}:
\ba
&&\epsilon_{\rm HB}=
\frac{3}{5}E_F^{(0)}u^{\frac{5}{3}}
\rho_0+V(u)\no
&&\;\;\;\;\;\;+u\rho_0(1-2x)^2S(u)
-\frac{\mu^2}{2}f_\pi^2\sin^2\theta
+2m_K^2f_\pi^2\sin^2\frac{\theta}{2}
+\mu u\rho_0x\no
&&\;\;\;\;\;\;-\mu
u\rho_0(1+x)\sin^2
\frac{\theta}{2}+(2a_1x+2a_2+4a_3)m_s
u\rho_0\sin^2\frac{\theta}{2}
-\frac{\mu^4}{12\pi^2} \; ,
\label{eHB}
\ea
where $E_F^{(0)}=(p_F^{(0)})^2/(2m_N)$ and
 $p_F^{(0)}=(3\pi^2\rho_0/2)^{1/3}$
are the Fermi energy and the Fermi momentum at normal nuclear
matter density, $x$ denotes the proton fraction $\rho_p=x\rho_B$,
and $u$ is defined by $\rho_B=u\rho_0$.\footnote{ To conform with
the expressions in the literature, we adopt the symbol $u$ here
instead of $c$ used in section 2.} $V(u)$ is the
charge-symmetric contribution of the nuclear interactions, while
$S(u)$ is the symmetry energy parameterized as \ba
S(u)=(2^{2/3}-1)\frac{3}{5}E_F^{(0)} (u^{2/3}-F(u))+S_0F(u)\, ,
\ea where $S_0\simeq 30~{\rm MeV}$ and we choose $F(u)=u$.
To concentrate on the effects of
kaon condensation on the quark-antiquark condensate,
we may ignore the first two terms in Eq. (\ref{eHB}),
which are independent of kaon condensation.
Furthermore, for simplicity, we ignore
the muon Fermi sea.
The energy of the charge-neutral ground state
for a given baryon density
is determined by extremizing
$\epsilon_{\rm HB}$ with respect
 to $x$, $\mu$ and $\theta$
\be
\frac{\partial
 \epsilon_{\rm HB}}{\partial x}=0,~\frac{\partial
 \epsilon_{\rm HB}}{\partial \mu}=0,~\frac{\partial
 \epsilon_{\rm HB}}{\partial \theta}=0\, .\label{3c}
\ee

We refer to Ref.~\cite{tpl94} for the explicit expressions
for these constraints.  The numerical solutions
for the above equations can be found
in Tables 3, 4 and 5 of Ref.~\cite{tpl94}.
{}Following the same procedure
as used in subsection 3.1, we obtain
\ba
&&\la \bar uu\ra =1-\frac{\sigma_N}{m_\pi^2 f_\pi^2}
[\frac{\partial
 \epsilon_{\rm HB}}{\partial m_N}
 +\frac{m_u}{\sigma_N}\frac{m_K^2}{m^\prime}
\frac{\partial
 \epsilon_{\rm HB}}{\partial m_K^2}] \no
&&\la \bar dd\ra =1-\frac{\sigma_N}{m_\pi^2 f_\pi^2}
\frac{\partial
 \epsilon_{\rm HB}}{\partial m_N} \, ,
\ea
where
\ba
&&\frac{\partial\epsilon_{\rm HB}}
{\partial m_N}=-\frac{1}{m_N}
u\rho_0 (1-2x)^2
 (2^{2/3}-1)\frac{3}{5} E_F^{(0)}(u^{2/3}-u)\label{kdd} \\
&&\frac{\partial
 \epsilon_{\rm HB}}{\partial m_K^2}=2f_\pi^2\sin^2
\frac{\theta}{2}\, .
\ea
Here $\sigma_N=m_qdm_N/dm_q$ with
$m_q=(m_u+m_d)/2$. We assume that
 $\sigma_N=\sigma_N^{(u)}$
($m_udm_N/dm_u$$)=\sigma_N^{(d)}$ ($m_d dm_N/dm_d)$.
It is now straightforward
to deduce the quark-antiquark condensate
in a kaon-condensed phase with the use of
$\epsilon_{\rm HB}$ along with the numerical
tables (Tables 3, 4 and 5) in Ref.~\cite{tpl94}.
The results are given in Table 1
for the representative values of input parameters,
$\sigma_N=30~{\rm MeV}$,
$a_1m_s=-67~{\rm MeV}$, $a_2m_s=134~{\rm MeV}$ and
$a_3m_s=(-134,-222, -310)~{\rm MeV}$.
Note that we have ignored the density dependence of
the quark-antiquark condensate
that is independent of kaon condensation.
Table 1 indicates that
the main conclusion of subsection 3.1
remains essentially unchanged
by the imposition of the $\beta$-equilibrium
and charge-neutrality conditions;
namely, $\frac{\la \bar uu\ra_{\rho_B}}
{\la \bar dd\ra_{\rho_B}}<1$
in kaon condensed phase.

  \begin{center}
  Table 1 : \parbox[t]{5.3in}{The ratios
  of quark-antiquark condensates
  in kaon condensed phase. Here ${\rm R}_u
=\frac{\la \bar
  uu\ra_{\rho_B}}{\la \bar uu\ra_{vac}}$
and ${\rm R}_d =\frac{\la \bar
  dd\ra_{\rho_B}}{\la \bar dd\ra_{vac}}$.
$a_3m_s$ and $\mu$ are given
  in units of MeV.}
  \end{center}
  $$
  \begin{array}{|r|r||r|r|r|}
  \hline
 a_3m_s&u &\mu&
{\rm R}_u &{\rm R}_d \\
  \hline
  \hline
  -134& 7.18&93.7&0.834&0.9998  \\ \hline
 -134& 7.68&73.5&0.822& 0.9999 \\ \hline\hline
 -222& 4.08&98.7&0.865&0.9997  \\ \hline
-222& 4.58&38&0.812& 0.9999\\ \hline\hline
-310& 2.92&86&0.875&0.9999  \\ \hline
 \end{array}
$$

We notice that $\la\bar dd\ra_{\rho_B}$ in Table 1
exhibits slight dependence on the kaon condensate,
whereas $\la\bar dd\ra_{\rho_B}$
in subsection 3.1 shows no such dependence.
This difference can be easily explained
by the fact that the symmetry energy $S(u)$, which is
responsible for the $\theta$-dependence of
$\la\bar dd\ra_{\rho_B}$,
cannot arise in tree-level or
one-loop-order calculations in the chiral quark model,
and therefore the effects subsumed in $S(u)$
are missing in the energy density calculated
in subsection 3.1.

 Before closing this subsection,
we discuss the effects of spontaneous isospin violation
summarized in Eq.~(\ref{isob}) on the nucleon mass.
In the simplest valence quark picture
in which the proton (neutron) contains two constituent $u$-quarks
and one constituent $d$-quark (one $d$-quark and two $u$-quarks),
we expect from Eq. (\ref{isob}) that the in-medium
proton and neutron masses in a kaon condensed phase
decreases with the density
faster than in a normal phase (without kaon condensation).
\footnote{A more detailed study on the nucleon mass and a
neutron-proton mass difference in kaon condensed matter will be
reported in a future publication. }
This feature is in qualitative agreement with the result
in Ref.~\cite{GS}.
The nucleon effective masses in a kaon condensed phase is
studied in the context of a relativistic mean-field model
in Ref.~\cite{GS}, and it was found that the nucleons
can have different effective masses in normal and kaon condensed
phases.

\section{Summary}

We have discussed $s$-wave $\Km$ condensation in the framework of
the chiral quark model, assuming that, in the density regime 
close to the critical density,  the relevant 
degrees of freedom are the  constituent 
quark degrees of freedom. 
We have primarily 
investigated the effects of charged kaon
condensation on the quark-antiquark condensate and we  
have found that a
$\Km$ condensate in quark matter suppresses 
the quark-antiquark
condensate for the $u$ quark, $\la\bar uu \ra$, but leaves
$\la\bar dd\ra$ unaffected in the lowest order approximation
adopted here.
This suggests the possibility that a partial chiral symmetry
restoration in the medium with a $\Km$ condensate may be flavor
dependent, {\it i.e.}, $\la \bar uu\ra/\la\bar dd\ra \ll 1$ for
increasing density. This raises an interesting question as to
whether or not the vector manifestation fixed point 
one finds in approaching
a chiral restoration point from normal Fermi liquid  remains 
intact if the chiral restoration point is approached from a kaon
condensed state with its distorted Fermi seas of quasiquarks.

\section*{Acknowledgments}
We are grateful for discussions with Gerry Brown, Chang-Hwan Lee
and Su Houng Lee. The work of YK, KK and FM is supported in part by the US
National Science Foundation, Grant Nos. PHY-0140214 and
PHY-0457014 and that of DPM by the BK21 project of the MOE, Korea.
Part of the work of MR was supported under Brain Pool Program of
Korea Research Foundation through KOFST, grant No. 051S-1-9.

\section*{Appendix A }
\setcounter{equation}{0}
\renewcommand{\theequation}{\mbox{A.\arabic{equation}}}
In this appendix, we state the power counting rules
for $\chi$QM.
Since we can treat the gluons perturbatively with
$\alpha_s\approx 0.28$~\cite{mg}, it suffices to focus on
the Goldstone bosons and quarks.

The most general vertex in $\chi$QM
in a cutoff regularization scheme takes the
form~\cite{mg},
\ba
(2\pi)^4\delta^4(\sum p_{i})(\frac{\pi}{f_\pi})^A
(\frac{\psi}{f_\pi\sqrt{\Lambda}})^B
(\frac{gG_\mu}{\Lambda})^C(\frac{p}{\Lambda})^D f_\pi^2\Lambda^2\ ,
\ea
 where for notational simplicity, we write
$\Lambda = \Lambda_{\chi SB}=4\pi f_\pi$.

\vspace{3mm}

As far as the Goldstone boson sector is concerned, this
counting rule is the same as the one used
in standard chiral perturbation theory(ChPT).
Including quarks is straightforward,
since the constituent quark mass, $M_f$,
can be considered small compared
to $\Lambda_{\chi SB}\sim 1$ GeV, i.e.
$M_f\sim p$ where $p$
is a typical momentum scale.
Each quark propagator contributes
$-1$ power of $p$, each Goldstone boson propagator contributes
$-2$ power of $p$, each derivative and quark mass in the
interaction terms contribute $+1$ power of $p$, and
each four-momentum integration contributes $+4$ powers of
$p$.

All the factors put together, the chiral index $D$ of a
given amplitude with $L$ loops, $I_{GB}$ internal meson lines,
$I_{Q}$ quark lines, $N^{GB}$ mesonic vertices and $N^{GBQ}$
meson-quark vertices is given by
\ba
D=4L -2I_{GB}-I_Q +\sum_n nN_n^{GB}+ \sum_d d N_d^{GBQ}.
\ea
For connected diagrams, we can use the topological relation
\ba
L=I_{GB}+I_Q - \sum_n (N_n^{GB}+N_n^{GBQ})+1
\ea
to get \ba D=2L +2 +I_Q +\sum_n (n-2)N_n^{GB}+ \sum_d (d-2)
N_d^{GBQ}.\label{ct}
 \ea


\begin{thebibliography}{99}

\bibitem{birse94}
M.C. Birse, J. Phys. G {\bf 20}  (1994) 1537.

\bibitem{kn}
D.B. Kaplan and A.E. Nelson, Phys. Lett. {\bf B175}, 57 (1986); {\bf 179} (1986)
409(E).

\bibitem{bkrt}
G.E. Brown, V. Thorsson, K. Kubodera and M. Rho,
Phys. Lett. {\bf B291} (1992) 355;
G.E. Brown, C.-H. Lee, M. Rho and V. Thorsson,
Nucl. Phys. {\bf A567} (1994) 937;
C.-H. Lee, G.E. Brown, D.-P. Min and M. Rho,
Nucl. Phys. {\bf A585} (1995) 401.

\bibitem{tpl94}
V. Thorsson, M. Prakash and J.M. Lattimer,
Nucl. Phys. {\bfA 572} (1994) 693.

\bibitem{Leeetal95}
C.-H. Lee, G.E. Brown, D.-P. Min and M. Rho,
Nucl. Phys. {\bf A585}  (1995) 401.

\bibitem{chlee}
C.-H. Lee, Phys. Rep. {\bf 275} (1996) 255.

\bibitem{mmttt} T. Maruyama, T.  Muto, T. Tatsumi, 
K. Tsushima and A. Thomas,
Nucl. Phys. {\bf A760} (2005) 319, and references therein.

\bibitem{mpp} D.P. Menezes, P.K. Panda and C. Providencia, 
Phys. Rev. {\bf C72} (2005) 035802. 

\bibitem{kw01}
N. Kaiser and W. Weise,
Phys. Lett. {\bf B 512} (2001) 283.

\bibitem{pjm02} T.-S. Park, H. Jung and D.-P. Min,
J. Korean Phys. Soc. {\bf 41}  (2002) 195.

\bibitem{BLKetal}  A. Bhattacharyya, S. K. Ghosh, S.C. Phatak and
  S. Raha,  Phys. Lett. {\bf B401}  (1997) 213; 
M. Lutz, Phys. Lett. {\bf B426} (1998) 12;
 E. E. Kolomeitsev and D. N. Voskresensky, Phys. Rev. {\bf C68}
  (2003) 015803. 

\bibitem{lsw92} M. Lutz, A. Steiner and W. Weise,
Phys. Lett. {\bf B 278} (1992) 29.

\bibitem{yabu93} H. Yabu, S. Nakamura, F. Myhrer 
and K. Kubodera,  Phys. Lett. {\bf B 315} (1993) 17; 
H. Yabu, S. Nakamura and K. Kubodera, 
Phys. Lett. {\bf B 317} (1993) 269; 
H. Yabu, F. Myhrer and K. Kubodera, 
Phys. Rev. {\bf D 50} (1994) 3549. 

\bibitem{WRW}
T. Waas, N. Kaiser and W. Weise,
Phys. Lett. {\bf B 379} (1996) 34;
T. Waas, M. Rho and W. Weise, Nucl. Phys. {\bf A 617}
   (1997) 449;
T. Waas and W. Weise, Nucl. Phys. {\bf A625} (1997) 287.

\bibitem{chp}
V. R. Pandharipande, C. J. Pethick and V. Thorsson,
Phys. Rev. Lett. {\bf 75} (1995) 4567;
J. Carlson, H. Heiselberg and V. R. Pandharipande,
  Phys. Rev. {\bf C 63} (2001) 017603.


\bibitem{bcpr} A. Barducci, R. Casalbuoni, G. Pettini and L. Ravagli,
Phys. Rev. {\bf D71} (2005) 016011.


\bibitem{hyper}
See for an example, A. Ramos, J. S.-Bielich and J. Wambach,
 Lect. Notes Phys. {\bf 578} (2001) 175.

\bibitem{Dka}
  Y.~Akaishi and T.~Yamazaki,
  Phys. Rev. {\bf C 65} (2002) 044005;
  A.~Dote, H.~Horiuchi, Y.~Akaishi and T.~Yamazaki,
  Phys. Rev. {\bf C 70} (2004) 044313;
  Y.~Akaishi, A.~Dote and T.~Yamazaki,
  Phys. Lett. {\bf B 613} (2005) 140;
  T.~Suzuki {\it et al.},
  Phys. Lett. {\bf B 597} (2004) 263;
  E.~Oset and H.~Toki,
  Phys. Rev. {\bf C 74}  (2006) 015207.
  M.~Agnello {\it et al.}
    [FINUDA Collaboration],
  Nucl. Phys. {\bf A  775} (2006) 35;
M.~Agnello {\it et al.}
  [FINUDA Collaboration],
Phys. Rev. Lett.  {\bf 94} (2005) 212303;
  V.~K.~Magas, E.~Oset, A.~Ramos and H.~Toki,
  Phys. Rev. {\bf C 74} (2006) 025206.
A. Gal, ``Overview of anti-K-Nuclear Quasi-Bound States.''
nucl-th/0610090.


\bibitem{LRS}  H.K. Lee, M. Rho and S.-J. Sin,
{Phys. Lett.} {\bf B348} (1995) 290.


\bibitem{ASY} E. Antonyan, J.A. Harvey, S. Jensen and D. Kutasov,
" NJL and QCD from string theory,'' hep-th/0604017;
O. Aharony, J. Sonnenschein and S. Yankielowicz, 
``A Holographic model of deconfinement and chiral symmetry
restoration,'' hep-th/0604161. 


\bibitem{MMT} D. Mateos, R. C. Myers and R. M. Thomson,
 Phys. Rev. Lett. {\bf 97} (2006) 091601.



 \bibitem{pseudogap}
H.-J. Lee, B.-Y. Park, M. Rho and V. Vento Nucl. Phys. {\bf A726} (2003)
69.


\bibitem{SSbaryon}
T. Sakai and S. Sugimoto, Prog. Theor. Phys. {\bf 113} (2005) 843;
K. Nawa, H. Suganuma and T. Kojo, `` Brane-induced skyrmions: Baryons
in holographic QCD,'' hep-th/0701007;
D. K. Hong, M. Rho, H.-U. Yee and P. Yi,  ``Chiral Dynamics of Baryons
from String Theory,'' hep-th/0701276;
H. Hata, T. Sakai, S. Sugimoto and S. Yamato,
 ``Baryons from instantons in holographic QCD,''  hep-th/0701280.


\bi{riska-brown}
D.O. Riska and G.E. Brown, Nucl. Phys. {\bf
A679} (2001) 577.

\bi{BR96}
G.E. Brown and M. Rho, Phys. Rep.  {\bf 269} (1996) 333.

\bi{BR01}
G.E. Brown and M. Rho, Phys. Rep. {\bf 363} (2002) 85.

\bibitem{BR04}
G.E. Brown and M. Rho, Phys. Rep. {\bf 398} (2004) 301.


\bibitem{BLR-star} G.E. Brown, C.-H. Lee and M. Rho,
Phys. Rev. {\bf C74} (2006) 024906.


\bibitem{mg}
A. Manohar and H. Georgi, Nucl. Phys. {\bf B234}, 189 (1984);
H. Georgi, {\it Weak Interactions and Modern Physics}
(Benjamin/Cummings, Reading. MA, 1984).


\bibitem{AHZ}R. Alkofer, S.T. Hong and I. Zahed,  J. Phys. {\bf G17} (1991) L59. 

\bibitem{BLR05} G.E. Brown, C.-H. Lee, H.-J. Park and M. Rho, 
Phys. Rev. Lett. {\bf 96} (2006) 062303. 


\bibitem{HY:PR}
M. Harada and K. Yamawaki, Phys. Repts. {\bf 381} (2003) 1.





\bibitem{MOW}
U.-G. Meissner, J. A. Oller and A. Wirzba,
    Ann. Phys. {\bf 297} (2002) 27; \\
L. Girlanda, A. Rusetsky and W. Weise, Ann. Phys. {\bf 312} (2004) 92.


\bibitem{cfg}
T. D. Cohen, R. J. Furnstahl and D. K. Griegel,
Phys. Rev. {\bf C45}  (1992) 1881.

\bibitem{LKetal} M. Lutz, B. Friman and C. Appel, Phys. Lett. {\bf
    B474} (2000) 7;
 N. Kaiser, S. Fritsch and W. Weise, Nucl. Phys. {\bf A697} (2002) 255.

\bibitem{sainio}
M. E. Sainio, PiN Newslett. {\bf 16} (2002) 138.


\bibitem{kubodera92}
K. Kubodera, J. Korean Phys. Soc. {\bf 26},  (1993) S171.


\bibitem{GS}
N. K. Glendenning and J. Schaffner-Bielich,
Phys. Rev. Lett. {\bf 81}  (1998) 4564.



\end{thebibliography}
\end{document}